\documentclass[a4paper,UKenglish,cleveref, autoref, thm-restate]{lipics-v2021}

\pdfoutput=1 %
\hideLIPIcs  %
\nolinenumbers

\newcommand{\Comment}[1]{\textsl{#1}}
\newcommand{\ouralg}{\textsc{HeiCut}}
\def\MdR{\ensuremath{\mathbb{R}}}

\DeclareMathOperator*{\argmax}{arg\,max}

\newtheorem{pruningrule}{Reduction Rule}
\newcommand{\set}[1]{\left\{ #1\right\}}

\usepackage{numprint} %
\usepackage{booktabs}
\usepackage{multirow}
\usepackage{array}
\usepackage{graphicx}
\usepackage{colortbl}
\usepackage{tabularx}
\usepackage{longtable}
\usepackage[usenames,dvipsnames]{xcolor}
\usepackage{tcolorbox} 
\definecolor{shade1}{gray}{0.95} %
\definecolor{shade2}{gray}{0.99} %
\usepackage{placeins}
\usepackage{tikz}
\usetikzlibrary{external}
\usepackage[ruled]{algorithm}
\usepackage[noend]{algpseudocode}
\usepackage[labelformat=simple]{subcaption}

\title{Near-Optimal Minimum Cuts in Hypergraphs at Scale} %

\author{Adil {Chhabra}}{Heidelberg University}{adil.chhabra@informatik.uni-heidelberg.de}{https://orcid.org/0009-0009-5726-9389}{}%
\author{Christian {Schulz}}{Heidelberg University}{christian.schulz@informatik.uni-heidelberg.de}{https://orcid.org/0000-0002-2823-3506}{}
\author{Bora {U\c{c}ar}}{CNRS and LIP (UMR5668 Université de Lyon - ENS de Lyon - UCBL - CNRS - Inria)}{bora.ucar@ens-lyon.fr}{https://orcid.org/0000-0002-4960-3545}{}
\author{Loris {Wilwert}}{Heidelberg University}{loris.wilwert@stud.uni-heidelberg.de}{https://orcid.org/0009-0003-3607-4992}{}

\authorrunning{A. Chhabra et. al.} %

\Copyright{Jane Open Access and Joan R. Public} %



\category{} %

\relatedversion{} %

\acknowledgements{We acknowledge support by DFG grant SCHU 2567/5-1. 
	Moreover, we would like to acknowledge Dagstuhl Seminar 23331 on Recent Trends in Graph Decomposition.
}

\EventEditors{John Q. Open and Joan R. Access}
\EventNoEds{2}
\EventLongTitle{42nd Conference on Very Important Topics (CVIT 2016)}
\EventShortTitle{ESA 2025}
\EventAcronym{CVIT}
\EventYear{2016}
\EventDate{December 24--27, 2016}
\EventLocation{Little Whinging, United Kingdom}
\EventLogo{}
\SeriesVolume{42}
\ArticleNo{23}

\begin{document}

\maketitle

\begin{abstract}
 The hypergraph minimum cut problem aims to partition its vertices into two blocks while minimizing the total weight of the cut hyperedges. This fundamental problem arises in network reliability, VLSI design, and community detection. We present \ouralg, a scalable algorithm for computing near-optimal minimum cuts in both unweighted and weighted hypergraphs. \ouralg~aggressively reduces the hypergraph size through a sequence of provably exact reductions that preserve the minimum cut, along with an optional heuristic contraction based on label propagation. It then solves a relaxed Binary Integer Linear Program (BIP) on the reduced hypergraph to compute a near-optimal minimum cut. Our extensive evaluation on over 500 real-world hypergraphs shows that \ouralg~computes the exact minimum cut in over 85\% of instances using our exact reductions alone, and offers the best solution quality across all instances. It solves over twice as many instances as the state-of-the-art within set computational limits, and is up to five \hbox{orders of magnitude faster.}
 
\end{abstract}

\section{Introduction}
\label{sec:intro}
Many phenomena in communication networks, database systems, circuit design, and computational biology involve interactions among more than two entities. Hypergraphs capture these higher-order relationships by allowing hyperedges -- generalized edges that can connect any number of vertices -- to span arbitrary subsets of vertices. Unlike traditional graphs, which model only pairwise interactions, hypergraphs represent multiway dependencies directly. This makes hypergraphs well-suited for representing group-level dependencies~\cite{doi:10.1126/science.aad9029}.

In this work, we tackle the minimum cut problem, which seeks to partition the vertices of a hypergraph into two non-empty sets such that the sum of the weights of the edges crossing the partition is minimized. A scalable solver for this problem has significant potential to support several applications that currently rely on graph-based solvers, including those in network reliability~\cite{kargernetworkreliability,Ramanathannetworkreliability}, VLSI design~\cite{KrishnamurthyVLSI}, clustering~\cite{HARTUV2000175,wucutclustering1993}, community detection~\cite{cai2005community}, and combinatorial optimization~\cite{padbergtsp}. 
For instance, in network reliability, the smallest edge cut is most likely to disconnect the network when all edges fail with equal probability. In VLSI design, minimizing interconnections between microprocessor blocks results in improved performance and cost reduction. In community detection, the absence of a small cut within a region of the graph can indicate the presence of a cluster. Thus, minimum cut algorithms are applied when recursively clustering graphs to maximize the minimum cut~\cite{cai2005community,HARTUV2000175}. While the graph minimum cut problem has been studied extensively -- leading to efficient combinatorial and randomized algorithms~\cite{haomincut1992, henzingerflowedgecon2020, kargermincut1996, nagamochiedgecon1992} and scalable solvers such as \textsc{VieCut}~\cite{viecut} -- the hypergraph setting remains more challenging. There is currently no practically scalable algorithm for the hypergraph minimum cut problem, despite its greater relevance in applications like VLSI design, where hypergraph cuts are of \hbox{greater practical interest~\cite{Lengauer1990}.}

Early approaches to the hypergraph minimum cut problem compute the minimum cut by solving a sequence of $(n - 1)$ $s$–$t$ cuts, using an ordering of the vertices in which the final two vertices, $s$ and $t$, define a minimum $s$–$t$ cut~\cite{kargermincut1996,makwongmincut2000}.
Recent work by Chekuri et al.~\cite{chekurihypergraphmincut} applied these ordering-based solvers after preprocessing the hypergraph. In particular, they introduce sparsification techniques that construct a \emph{$k$-trimmer certificate} -- a sub-hypergraph that preserves all local connectivities up to size $k$. This yields an algorithm for computing minimum cuts in unweighted hypergraphs, running in $O(p + \lambda n^2)$ time, where $\lambda$ is the minimum cut, $n$ is the number of vertices, and $p$ is the number of times vertices appear in hyperedges. 
This approach has non-linear running time, and does not extend to weighted instances, leaving open the challenge of 
designing a general-purpose solver for computing minimum cuts in large hypergraphs.

\textbf{Our contribution.} In this work, we present \ouralg, a near-optimal and highly efficient solver for the hypergraph minimum cut problem that supports both weighted and unweighted instances. Our algorithm combines novel, hypergraph-specific exact reduction rules with generalized adaptations of effective graph-based reductions from \textsc{VieCut}~\cite{viecut}. These reductions enable aggressive contractions while provably preserving the minimum cut, substantially reducing instance size. When no further reductions are possible, we solve the remaining instance using a near-optimal relaxed Binary Integer Programming (BIP) formulation. Additionally, we offer an optional heuristic reduction based on clustering with label propagation, which is more memory-efficient while still achieving comparable solution quality in most instances. To evaluate performance, we also introduce a new dataset of synthetic $k$-core hypergraphs with non-trivial minimum cuts -- designed to serve as a benchmark for future work on \hbox{this problem.} 

Our extensive experiments on both real-world and synthetic instances demonstrate that \ouralg~consistently computes near-exact solutions and sets a new state-of-the-art in terms of scalability, efficiency, and robustness across diverse hypergraphs, being up to five orders of magnitude faster than the current state-of-the-art. Although our relaxed BIP solver has non-linear complexity, our reduction rules are so effective that in over 85\% of real-world instances, they suffice to uncover the exact minimum cut without invoking the BIP solver. This allows \ouralg~to achieve near-linear \hbox{runtime in practice.}

\section{Preliminaries}
\subsection{Basic Concepts}
An \textit{undirected hypergraph} $H=(V,E)$ is defined as a set of vertices $V$ and a
multiset of hyperedges $E$, where each hyperedge~$e\in E$ is a subset of vertices.
The vertices that compose a hyperedge~$e\in E$ are called the~\emph{pins} of~$e$. We denote~$n = |V|$ as the number of vertices and~$m = |E|$ as the number of hyperedges.
The~\emph{size} of a hyperedge~$e$ is~$|e|$, which is the number of pins contained in~$e$.
We denote~$p := \sum_{e \in E} |e|$ as the total number of pins in the hypergraph~$H$. 
Let~$c: V \rightarrow \MdR_{\geq 0}$ be a vertex-weight function and let~$\omega: E \rightarrow \MdR_{\geq 0}$ be a hyperedge-weight function. The functions~$c$ and~$\omega$ are generalized to sets, such that~$c(V') := \sum_{v \in V'} c(v)$ for~$V' \subseteq V$ and~$w(E') := \sum_{e \in E'} \omega(e)$ for~$E' \subseteq E$. A \emph{connected path}~$P$ in~$H$ is a (finite) sequence of~$s$ pairwise distinct edges~$(e_1, ..., e_s)$ such that $e_i \cap e_{i+1} \neq \emptyset~\forall~1\leq i < s$.

We assume hyperedges to be sets rather than multisets; that is, a vertex can be contained in a hyperedge only \emph{once}, while multiple edges can contain the same set of vertices. A hyperedge~$e \in E$ is~\emph{incident} to a vertex~$v \in V$ if~$v \in e$. Two vertices~$u, v$ are said to be~\emph{adjacent} if at least one hyperedge is incident to both of them. For a vertex~$v \in V$, we define~$I(v) := \{e \in E : v \in e \}$ as the set of incident hyperedges to the vertex~$v$ and generalize it to sets, such that~$I(V') := \bigcup_{v \in V'} I(v)$. We define~$N(v) := \{u \in V~\backslash~\{v\} : I(u) \cap I(v) \neq \emptyset \}$ as the~\emph{neighborhood} of the vertex~$v$. Let~$d(v) := |I(v)|$ be the~\emph{degree} of~$v$ and~$d_{\omega}(v) := \omega(I(v))$ be the~\emph{weighted degree} of~$v$. We denote~$\delta := \min_{v \in V}\{d(v)\}$ as the minimum (unweighted) degree, and~$\Delta := \max_{v \in V}\{d(v)\}$ as the maximum (unweighted) degree of~$H$. The weighted counterparts are denoted by~$\delta_{\omega}$ and~$\Delta_{\omega}$.

A~\emph{contraction} of two vertices~$u, v \in V$ in a hypergraph~$H$ consists of merging the vertices~$u$ and~$v$ to a new vertex~$w$ so that~$c(w) = c(u) + c(v)$ and~$I(w) = I(u) \cup I(v)$. 
A~\emph{contraction} of a hyperedge~$e = {v_1, \dots, v_k} \in E$ replaces the vertices~$v_1, \dots, v_k$ with a new vertex~$w$ such that~$c(w) = \sum_{i=1}^k c(v_i)$ and~$I(w) = \bigcup_{i=1}^k I(v_i) \setminus {e}$. The hyperedge~$e$ is removed from~$E$, and all occurrences of~$v_1, \dots, v_k$ in other hyperedges are replaced by~$w$.

\subparagraph{Hypergraph Minimum Cut.} The~\emph{hypergraph minimum cut problem} consists of cutting the vertex set~$V$ of the hypergraph~$H$ into two non-empty~\emph{blocks}~$A$ and~$V~\backslash~A$ with the objective of minimizing the~\emph{value/capacity} of the cut~$(A,~V~\backslash~A)$, which is the sum of the weights of all edges running between the two blocks:
\begin{equation}
    \begin{aligned}
	\lambda (H) := \min_{\emptyset \neq A \subsetneq V} \left\{ \sum_{e \in E} \{ \omega(e) : e \cap A \neq \emptyset \wedge e \cap V~\backslash~A \neq \emptyset \} \right\}
    \end{aligned}
\end{equation}
For unweighted hypergraphs, $\omega(e) = 1~\forall~e\in E$. We denote~$\lambda(H)$ (or just~$\lambda$ if the context is clear) as the value of a minimum cut of the hypergraph~$H$. During the execution of our algorithm, $\hat{\lambda}(H)$ (or just~$\hat{\lambda}$) represents the smallest upper bound of the minimum cut value~$\lambda(H)$ found so far. The~\hbox{\emph{trivial cut}} is defined as~$(\{v\},~V~\backslash~\{v\})$, where~$v\in V$ is the vertex with the smallest weighted vertex degree, i.e.~$d_{\omega}(v)=\delta_{\omega}$.

\subsection{Related Work}
\label{sec:related_work}
For graphs, a wide range of efficient algorithms have been developed for computing the minimum cut~\cite{haomincut1992,henzingerflowedgecon2020,kargermincut1996,nagamochiedgecon1992}. Traditional approaches relied on solving $(n - 1)$ $s$-$t$ minimum cut problems via maximum flow computations, where a~\emph{minimum s-t cut} is defined as the value of a minimum cut separating the vertices~$s, t \in V$. Nagamochi and Ibaraki~\cite{nagamochiedgecon1992} introduced the maximum adjacency (MA) ordering, a greedy strategy that avoids explicit flow computations. In this scheme, the final two vertices in the MA ordering are guaranteed to define a minimum $s$-$t$ cut, and thus it is possible to compute the $(n-1)$ $s$-$t$ minimum cuts without computing network flows. This idea, along with Karger's randomized contraction algorithm~\cite{kargermincut1996} and its later refinements using cut sparsification and tree packings \cite{karger2000mincut}, forms the basis of many fast algorithms for the graph minimum cut problem. More recently, Henzinger et al.~\cite{viecut} proposed \textsc{VieCut}, a linear-time algorithm that represents the current state-of-the-art for computing graph minimum cuts. \textsc{VieCut} utilizes a fast, parallel \emph{inexact} minimum cut method to compute a good approximate bound for the problem. Using this bound, \textsc{VieCut} performs aggressive graph reductions through safe edge contractions with advanced data structures and parallel contraction routines. Ultimately, \textsc{VieCut} runs an \emph{exact} minimum cut solver on the reduced graph based on the algorithms of \hbox{Nagamochi et al.~\cite{nagamochiedgecon1992,nagamochimincut1994}.} 

Extending the minimum cut problem to hypergraphs presents additional challenges. A naïve approach models hypergraphs as their corresponding bipartite graph representation, and then performs $(n - 1)$ flow computations. %
However, Queyranne~\cite{Queyranne1998} showed that ordering schemes like the MA ordering apply to arbitrary symmetric submodular functions, and therefore can be extended to hypergraphs.
This result enables vertex-ordering based deterministic algorithms for computing exact minimum cuts in both weighted and unweighted hypergraphs such as those developed by Klimmek and Wagner~\cite{Klimmek1996} and by Mak and Wong~\cite{makwongmincut2000}.

Building on these ordering approaches, Chekuri et al.~\cite{chekurihypergraphmincut} introduced sparsification techniques to accelerate hypergraph minimum cut computation. Specifically, they proposed a preprocessing phase that simplifies the hypergraph~$H$ to a \hbox{\emph{$k$-trimmed certificate}}~$H_k$ with~$O(kn)$ hyperedges, which is then passed to a vertex-ordering based exact solver. The~$k$-trimmed certificate~$H_k$ preserves all local connectivities in~$H$ up to~$k$, i.e.,~$\lambda(H_k, s, t) \geq \min{k, \lambda(H, s, t)}$ for all~$s, t \in V$, where~$\lambda(H, s, t)$ denotes the minimum cut between $s$ and $t$. The minimum cut value is preserved only if~$\lambda(H) = \lambda(H_k) < k$. Thus, the algorithm iteratively doubles~$k$, starting from~$k = 2$, and computes~$\lambda(H_k)$ using the exact solver in each iteration. If~$\lambda(H_k) < k$, then~$\lambda(H) = \lambda(H_k)$ and the algorithm stops; otherwise,~$k$ is doubled. The algorithm runs in~$O(p + \lambda n^2)$ time and applies only to unweighted hypergraphs. 

\section{\textsc{HeiCut}: Near-Exact Hypergraph Minimum Cuts at Scale}
Computing exact minimum cuts in hypergraphs is a fundamental yet computationally demanding problem, especially at scale. Directly applying exact algorithms to large hypergraphs is often infeasible due to their size and complexity. In this section, we present \ouralg, a scalable algorithm for computing near-optimal minimum cuts in large hypergraphs. The central idea behind \ouralg~is to aggressively reduce the hypergraph size through a sequence of provably exact reductions that preserve the minimum cut. Once the instance is small enough, we compute the minimum cut using our near-optimal relaxed Binary Integer Linear Program (BIP) formulation (henceforth, \textsc{Relaxed-BIP}). 
To further enhance reduction effectiveness, \ouralg~also supports an optional heuristic contraction step based on label propagation. A pseudocode depicting the overall approach is shown in Algorithm~\ref{alg:ouralg}.

We begin this section by detailing the exact reduction rules that shrink the hypergraph without affecting the minimum cut. We then introduce our BIP formulation for computing the minimum cut on the reduced instance. Finally, we describe the optional label propagation heuristic, which clusters vertices to enable additional contractions.

\begin{algorithm}[t]
	\caption{\ouralg: Near-Optimal Hypergraph Minimum Cut}
	\label{alg:ouralg}
	\begin{algorithmic}[1]
		\State \textbf{Input:} Hypergraph $H = (V, E)$, \textcolor{blue}{$useLP$} \Comment{\textcolor{gray}{optional heuristic LP reduction}}
		\State \textbf{Output:} Minimum cut value $\lambda$
		
		\State Initialize upper bound $\hat{\lambda} \gets \min_{v \in V} d_\omega(v)$

		\While{$|V| > n_o$ or not reducing further}
            \If{\textcolor{blue}{$useLP$}} 
    			\State $H \gets$ \Call{LabelPropagationContraction}{$H$}
    		\EndIf
			\For{\textbf{each} exact reduction rule $r$ in fixed order}
				\State $H \gets$ \Call{ApplyExactReduction}{$H, r, \hat{\lambda}$}
				\State Update $\hat{\lambda} \gets \min(\hat{\lambda}, \delta_\omega(H))$
			\EndFor
        \EndWhile
        
        \If{$|E| = 0$}
        	\State $\lambda \gets 0$ \Comment{\textcolor{gray}{already fully reduced}}
        	\ElsIf{$|V| = 1$} \State $\lambda \gets \hat{\lambda}$
        	\Else \State $\lambda \gets$ \Call{SolveRelaxedBIP}{$H$}
        \EndIf
     
		\State \Return $\lambda$
	\end{algorithmic}
\end{algorithm}

\subsection{Exact Reductions}
\label{subsec:reductions}
Before solving the hypergraph minimum cut problem using \textsc{Relaxed-BIP}, we apply a sequence of contraction rounds guided by a set of seven exact reduction rules. These rules identify structures within the hypergraph whose contraction provably preserves the value of the minimum cut, ensuring that any minimum cut of the reduced hypergraph remains valid on the input hypergraph. The reduction process terminates once the size of the hypergraph falls below a predefined vertex threshold, or no further reductions occur between rounds.

Each reduction round considers each of the seven rules in a fixed order, with each rule applied at most once per round. If a rule is applicable, it is immediately applied, contracting the hypergraph, before proceeding to the next rule. After each round, we check if the stopping criteria are met. If the hypergraph is reduced to a single vertex or a set of isolated vertices, the minimum cut value has been determined, and \textsc{Relaxed-BIP} is not required. 

To guarantee that contractions do not increase the true minimum cut value, we maintain an upper bound $\hat{\lambda}$ on the minimum cut, initialized with the minimum weighted vertex degree.
Our reduction rules are designed to contract only those components that cannot participate in any cut of value less than $\hat{\lambda}$. That is, if such a cut exists, the contracted components are entirely contained within one side of the partition and thus do not cross the cut. Consequently, these contractions preserve the correctness of the solution.

After each contraction within a round, we update the upper bound~$\hat{\lambda}$ if the minimum vertex degree of the reduced hypergraph decreases, and use this tighter bound when evaluating the subsequent reduction rules. Maintaining and tightening this bound enables more aggressive reductions, as some rules become applicable only under smaller values of $\hat{\lambda}$.

\begin{figure}[t]
    \centering
    \includegraphics[width=0.9\linewidth]{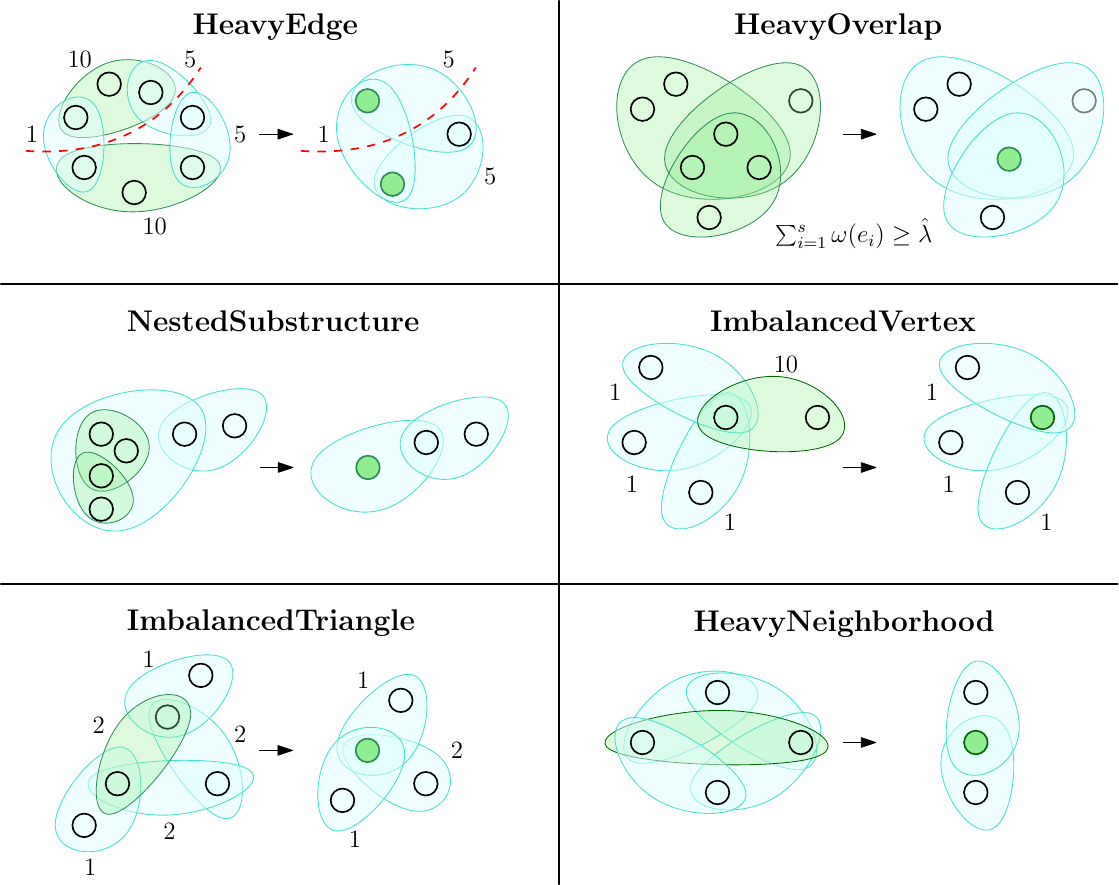}
    \caption{Depictions of our various reduction rules, excluding Rule 1, which is self-evident.}
    \label{fig:redrules}
\end{figure}

We now describe the set of exact reduction rules used in our contraction pipeline, depicted in Figure~\ref{fig:redrules}.~The rules are applied in the order in which they are presented, prioritizing those with low computational overhead that are expected to yield significant reductions in practice.
The \texttt{HeavyEdge} reduction (Reduction Rule~\ref{pr:heavy_edges}) is a technique originally introduced by Padberg and Rinaldi~\cite{Padberg1990} for graphs and later employed in the \textsc{VieCut} algorithm~\cite{viecut}. We generalized it to hypergraphs where hyperedges can be of arbitrary size. In contrast, Rules~\ref{pr:viecut_2}, \ref{pr:viecut_3}, and \ref{pr:viecut_4} are non-generalized reduction rules applicable only to standard graph edges; these are directly inherited from the \textsc{VieCut} framework and are only applied to hyperedges of size two. These graph-based rules are applied last, as contractions from earlier rules may increase the number of hyperedges of size two, thereby expanding their applicability. Modifications to the application of these graph-based reduction rules are discussed following the complete list of reductions. Finally, Rules~\ref{pr:heavy_overlap} and~\ref{pr:strictly_nested_isolated_substructures} are novel hypergraph-specific reduction rules that exploit structural properties unique to hyperedges and have no analog in the graph setting.

\begin{pruningrule}[\textbf{Singleton}]~\label{pr:trivial_edges}
	Remove any hyperedge~$e \in E$ where~$|e| = 1$ or $\omega(e) = 0$.\\
	\textnormal{\textbf{Proof}}: This pruning rule holds trivially since a hyperedge $e$ with $|e| = 1$ cannot be cut, and if it has weight $\omega(e) = 0$, it can not contribute to the cut value. \\
	\textnormal{\textbf{Complexity}}:~$O(m)$ when iterating once over all hyperedges and removing if applicable.
\end{pruningrule}

\begin{pruningrule}[\textbf{HeavyEdge}]~\label{pr:heavy_edges}
	Contract hyperedges~$e \in E$ where~$|e| \geq 2$ and $\omega(e) \geq \hat{\lambda}$. \\
	\textnormal{\textbf{Proof}}: Let $\hat{\lambda}$ be the lowest upper bound of the minimum cut. Initially, we use the minimum degree as the upper bound, i.e., $\hat{\lambda} = \delta_{\omega}$, which is a trivial cut. Now, assume that a hyperedge $e$ with $|e| \geq 2$ and weight $\omega(e) \geq \hat{\lambda}$ crosses a minimum cut $\lambda$. Then,~$\lambda \geq \omega(e) \geq \hat{\lambda}$. Since $\lambda \geq \hat{\lambda}$, and $\hat{\lambda}$ is an upper bound, the minimum cut must be equal to the upper bound we had already discovered, i.e., $\lambda = \hat{\lambda}$. Thus, any cut that separates the vertices of $e$ cannot yield an improvement over the current bound and thus $e$ can be safely contracted. \\
	\textnormal{\textbf{Complexity}}:~$O(n+m)$ when iterating over all hyperedges and contracting them if applicable.
\end{pruningrule}

\begin{pruningrule}[\textbf{HeavyOverlap}]~\label{pr:heavy_overlap}
	For any overlap of~$s \geq 2$ hyperedges~$\bigcap_{i=1}^{s} e_i$ where \hbox{$\left|\bigcap_{i=1}^{s} e_i \right| \geq 2$ and $\sum_{i=1}^{s} \omega(e_i) \geq \hat{\lambda}$}, contract $\bigcap_{i=1}^{s} e_i$ into a single vertex. \\
	\textnormal{\textbf{Proof}}: Assume that some overlap~$\bigcap_{i=1}^{s} e_i$ of hyperedges is cut in the minimum cut $\lambda$. This implies that all $s$ hyperedges $e_i$ that share the overlap are cut in the minimum cut, with each contributing a weight of $\omega(e_i)$ to the cut. Then~$\lambda \geq \sum_{i=1}^{s} \omega(e_i) \geq \hat{\lambda}.$ Similarly to the Reduction Rule 2 proof, it must then hold that the minimum cut is equal to the upper bound that we have already discovered, i.e. $\lambda = \hat{\lambda}$. Therefore, cutting the overlap does not improve the cut value, and thus we can safely contract the overlap. \\
	\textnormal{\textbf{Complexity}}:~$O(n + \sum_{e\in E} |e|^2)$ since we visit each pin in the incident hyperedges of every vertex~$v \in V$ with~$d_{\omega}(v) \geq \hat{\lambda}$.
\end{pruningrule}

\begin{pruningrule}[\textbf{NestedSubstructure}]~\label{pr:strictly_nested_isolated_substructures}
	For a hyperedge~$e \in E$, contract any strictly nested substructure~$\bigcup_{i=1}^s e_i \subsetneq e$ for which there exists no path~$P$ starting in the substructure and leaving $e$, such that every hyperedge $e' \in P$ satisfies $e \not\subseteq e'$.\\
	\textnormal{\textbf{Proof}}: Let~$e \in E$ be a hyperedge and~$U = \bigcup_{i=1}^s e_i$ a substructure such that each~$e_i \subseteq E$ and~$e_i \subsetneq e$, i.e., all vertices of $U$ are contained strictly within~$e$, and all~$e_i$ are also subsets of~$e$. Moreover, assume that there exists no path~$P$ from any vertex in~$U$ to any vertex outside of~$e$, such that every hyperedge~$e' \in P$ satisfies $e \not\subseteq e'$. This implies that all connectivity from $U$ to $V \setminus U$ must go through~$e$.
    Now, consider an arbitrary minimum cut $(A, B)$ in the hypergraph. Since~$e$ is the only hyperedge through which vertices in~$U$ connect to the rest of the hypergraph, and~$e$ contains all of~$U$, the substructure~$U$ cannot be separated from the rest of the hypergraph except via~$e$. Thus, the side of the cut to which the vertices of~$U$ belong does not affect whether or not $e$ is cut. Therefore, we can safely assign all of $U$ to either side of the cut without changing the cut value.
    Thus, the substructure~$U$ can be contracted into a single vertex without affecting \hbox{the minimum cut.} \\
	\textnormal{\textbf{Complexity}}:~$O(n+ m + \Delta \cdot p + \sum_{e\in E} \sum_{e' \in E \wedge e' \subsetneq e} |e'|)$ as each parent hyperedge $e$ iterates over all its pins and their incident hyperedges to detect strictly nested substructures, then performs traversals to verify whether pins can escape through other incident edges $e'$. 
\end{pruningrule}

\begin{pruningrule}[\textbf{ImbalancedVertex}]~\label{pr:viecut_2}
	Contract any hyperedge~$e_{uv} = \{u, v\} \in E$ where~$d_{\omega}(u) < 2 \omega(e_{uv})$ or $d_{\omega}(v) < 2 \omega(e_{uv})$  \\
	\textnormal{\textbf{Proof}}: We only apply this rule to hyperedges of size 2, which have the same properties as edges of regular graphs. The rule has been proven for $|e_{uv}| = 2$ in~\cite{Padberg1990}. \\
	\textnormal{\textbf{Complexity}}:~$O(n + m)$ when iterating over all hyperedges and contracting them if applicable.
\end{pruningrule}

\begin{figure}
    \centering
    \includegraphics[width=0.5\linewidth]{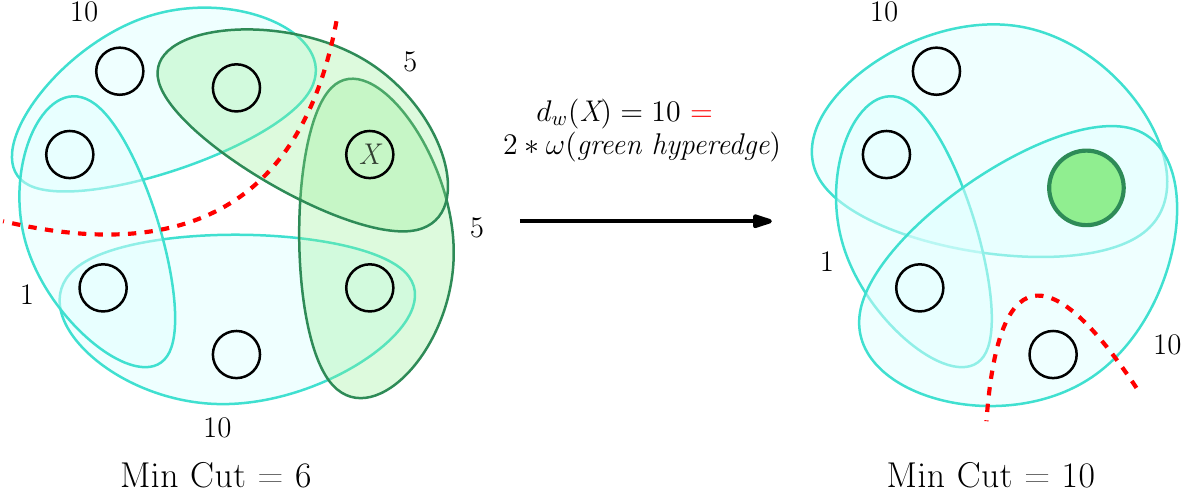}
    \caption{A sample hypergraph with a minimum cut value of six. When using a non-strict inequality in Reduction Rule~\ref{pr:viecut_2}, we contract the green hyperedges simultaneously, which increases the minimum cut in the contracted hypergraph.}
    \label{fig:pruningrule5problem}
\end{figure}

\begin{pruningrule}[\textbf{ImbalancedTriangle}]~\label{pr:viecut_3}
	Contract any hyperedge~$e_{uv} = \{u, v\} \in E$ where~$\exists w \in V: e_{uw} = \{u, w\} \in E$ and $e_{vw} = \{v, w\} \in E$ with~$d_\omega{(u)} \leq 2(\omega(e_{uv}) + \omega(e_{uw}))$ or $d_\omega{(v)} \leq 2(\omega(e_{uv}) + \omega(e_{vw}))$.  \\
	\textnormal{\textbf{Proof}}: Similar to Reduction Rule~\ref{pr:viecut_2}, this has been proven in~\cite{Padberg1990}. The intuition is that a hyperedge~$e_{uv}$ never crosses a minimum cut, because then~$e_{uw}$ or~$e_{vw}$ also crosses this minimum cut and shifting either~$u$ or~$v$ to the other block would decrease the cut value. \\
	\textnormal{\textbf{Complexity}}:~$O(n+m)$ when marking the vertices and thus not looking at all possibilities.
\end{pruningrule}

\begin{pruningrule}[\textbf{HeavyNeighborhood}]~\label{pr:viecut_4}
	Contract any hyperedge~$e_{uv} = \{u, v\} \in E$ where~$\omega(e_{uv}) + \sum_{w \in U} \min\{\omega(e_{uw}), \omega(e_{vw})\} \geq \hat{\lambda}$ with~$U := \{ w \in V : e_{uw} = \{u, w\} \in E \wedge e_{vw} = \{v, w\} \in E\}$. \\
	\textnormal{\textbf{Proof}}: As for Reduction Rule~\ref{pr:viecut_2} and~\ref{pr:viecut_3}, this has been proven for $|e_{uv}| = 2$ in~\cite{Padberg1990}. The key idea is that if we assume that the hyperedge~$e_{uv}$ crosses a minimum cut $\lambda$, then~$e_{uw}$ or~$e_{vw}$ also crosses this minimum cut for every~$w  \in U$, i.e.~$\lambda \geq \sum_{w \in U} \min\{\omega(e_{uw}), \omega(e_{vw})\} \geq \hat{\lambda}$. Then, as in the proof of Reduction Rules~\ref{pr:heavy_edges} and~\ref{pr:heavy_overlap}, the minimum cut is equal to the upper bound we have already discovered, that is, $\lambda = \hat{\lambda}$. \\
	\textnormal{\textbf{Complexity}}:~$O(n+m)$ when marking the vertices and thus not looking at all possibilities.
\end{pruningrule}

Reduction Rules~\ref{pr:viecut_2} and \ref{pr:viecut_3} are adapted slightly from their original formulations to ensure correctness in the hypergraph setting. 
We modify the inequality in Rule~\ref{pr:viecut_2} to be strict. This adjustment is necessary because, unlike the original formulation by Padberg and Rinaldi~\cite{Padberg1990}, a non-strict inequality may lead to erroneous contractions. Specifically, if two hyperedges $e_{uv}$ and $e_{vw}$ have equal weight and share a common pin $v$, both may be independently selected for contraction. This implicitly assumes that $v$ can be shifted to opposite sides of two different cuts simultaneously, which results in an invalid contraction that destroys the minimum cut. A counterexample illustrating this issue is provided in Figure~\ref{fig:pruningrule5problem}.

For Rule~\ref{pr:viecut_3}, the inequality remains non-strict, consistent with the implementation used in \textsc{VieCut}, where each vertex is explicitly marked to participate in at most one contraction per round. Additionally, we note that \textsc{VieCut} stipulates that its reduction rules apply only if the edge $e_{uv}$ is not the sole edge incident to its endpoints $u$ and $v$. This restriction prevents contractions that may eliminate non-trivial minimum cuts. However, in our setting, we find this condition to be unnecessary. As shown by Padberg and Rinaldi~\cite{Padberg1990}, only trivial cuts may be affected in such cases, and we already account for this by maintaining the minimum weighted vertex degree $\delta_{\omega}$ and updating the upper bound $\hat{\lambda}$ whenever $\delta_{\omega} < \hat{\lambda}$.

\subsection{Binary Integer Linear Program}
To compute a minimum cut on the reduced hypergraph instance, we formulate the hypergraph minimum cut problem as a Binary Integer Linear Program. For each vertex $v \in V$, we use a binary variable $x_v \in {0, 1}$ indicating the block assignment of the vertex. 
Additionally, for each hyperedge $e \in E$, we define a binary variable $y_e \in {0, 1}$ that denotes whether $e$ is cut. This results in a total of $n + m$ binary variables. The objective is to minimize the total weight of the hyperedges that are cut, leading to the following objective function:
\begin{equation} 
    \begin{aligned}
	\min{\sum_{e \in E} \omega(e) \cdot y_e}
    \end{aligned}
\end{equation}
We enforce two types of constraints: cut non-triviality and hyperedge-cut indicators. 
\begin{equation}
\label{bip-constraints}
\begin{aligned}
&\textnormal{(a)} \sum_{v \in V} x_v \geq 1 \quad \textnormal{(b)} \quad \sum_{v \in V} x_v \leq n - 1 \\
&\textnormal{(c)}~\forall e \in E,\ \forall (u, v) \in e \times e,\ u \neq v:\quad y_e \geq x_u - x_v
\end{aligned}
\end{equation}

The first two constraints in Equation~\ref{bip-constraints} ((a) and (b)) ensure both blocks are non-empty.
The third constraint (c) ensures that if any two vertices $u$ and $v$ in a hyperedge $e$ are assigned to different blocks ($x_u \neq x_v$), then $y_e = 1$, marking $e$ as cut. If all vertex assignments in $e$ are the same, the constraint does not force $y_e$ to be 1. This allows solutions where $y_e = 1$ even if all vertices in $e$ belong to the same block.
However, such solutions are clearly suboptimal.

In our algorithm, we solve a relaxed version of the BIP, allowing variables to take floating-point values near the binary domain. The solution is then rounded to the nearest integer to obtain a feasible binary result. This improves scalability, as solving the exact integer BIP is computationally expensive. Although the approach is inexact, the practical difference is negligible (see Section~\ref{subsec:real}). Thus, the relaxed BIP yields a near-optimal minimum cut.

\subsection{Heuristic Reduction via Label Propagation}
\ouralg~optionally applies a heuristic reduction based on clustering, inspired by \textsc{VieCut}. We contract dense vertex clusters before each round of the exact reductions described in Section~\ref{subsec:reductions}. 
The intuition is that vertices within a densely-connected cluster are unlikely to be split by a minimum cut, as this incurs a large cut cost. Thus, contracting such clusters early aggressively reduces the hypergraph size with minimal impact on cut quality.
While this reduction is inexact and lacks optimality guarantees, evidence from \textsc{VieCut} shows that solution quality remains high, often matching the exact cut even on large, \hbox{complex instances.}

We compute the clustering using label propagation, originally proposed for graphs~\cite{Raghavan_2007}, 
and
later extended to hypergraphs~\cite{Henne2015_1000063440}.
Each vertex starts in its own singleton cluster with a unique label corresponding to its vertex ID. The algorithm proceeds iteratively: per iteration, the vertices are processed in a fixed random order and assigned the label of the cluster to which they are most strongly connected, breaking ties uniformly at random. More formally, let $L$ be the current set of labels and $C_l$ the set of vertices assigned label $l \in L$. Each vertex $v \in V$ is reassigned to the label maximizing the following score: 
\begin{equation} \label{eq:absorptionclusteringusingpins}
\begin{aligned}
    \argmax_{l \in L} \left\{ \sum_{e \in I(v)} |e \cap C_l| \cdot \frac{\omega(e)}{|e| - 1}\right\}
\end{aligned}
\end{equation} 

This method, known as absorption clustering using pins, is equivalent to graph-based label propagation on the clique-expanded hypergraph, where each hyperedge $e \in E$ is replaced with a complete clique among its pins, and each resulting graph edge is weighted $\frac{\omega(e)}{|e| - 1}$. 
This model is widely used in hypergraph partitioning tools such as \textsc{hMetis}~\cite{DBLP:journals/tvlsi/KarypisAKS99}, \textsc{PaToH}~\cite{DBLP:journals/tpds/CatalyurekA99} and \textsc{Mt-KaHyPar}~\cite{mtkahypar}. 
The algorithm runs in $O(n + \sum_{e \in E} |e|^2)$ time, as each vertex considers clusters of all pins in its incident hyperedges. Despite this, it is efficient in practice and typically converges in a few iterations, with rapid convergence shown by Kothapalli et al.~\cite{DBLP:conf/icdcn/KothapalliPS13}. 

\section{Experimental Evaluation}
We evaluate the performance, scalability, and solution quality of \ouralg~on a diverse set of real-world and synthetic hypergraphs by answering three key questions:
\begin{itemize}
	\item \textbf{RQ1:} How effective are the proposed exact reduction rules in shrinking hypergraphs while preserving the minimum cut?
	
	\item \textbf{RQ2:} How does \ouralg~compare to the state-of-the-art hypergraph minimum cut algorithm in terms of solution quality, runtime and memory usage?
	
	\item \textbf{RQ3:} What is the trade-off between accuracy and efficiency when enabling heuristic reduction via label propagation?
\end{itemize}

We first describe our baselines, setup, instances, and evaluation methodology, then address our research questions by analyzing the impact of reductions on~\ouralg's cut quality (Section~\ref{subsec:impact_red}) and comparing baselines on real-world and \hbox{synthetic datasets (Sections~\ref{subsec:real} and~\ref{subsec:syn}).}

\subparagraph{Baselines.} We compare \ouralg~against two baselines: (i) 
the current state-of-the-art minimum cut solver for \emph{unweighted hypergraphs}~\cite{chekurihypergraphmincut}, which we refer to as \textsc{Trimmer}, and (ii) \textsc{Relaxed-BIP} as a standalone solver. Notably, \textsc{Trimmer} is an exact solver, while \ouralg~and \textsc{Relaxed-BIP} are near-optimal, but inexact. To the best of our knowledge, no implementation of the \textsc{Trimmer} algorithm is publicly available. 
Thus, we implemented \textsc{Trimmer} ourselves based on its description. The underlying exact algorithm in our implementation of \textsc{Trimmer} uses the MA ordering~\cite{chekurihypergraphmincut}. A pseudocode of our implementation can be found in Appendix~\ref{appendix:trimmer}. \textsc{Trimmer} is evaluated only on unweighted instances, as it does not solve the weighted case. For the same reason, \textsc{Trimmer} cannot be used as a solver in place of \textsc{Relaxed-BIP} in \ouralg, as our reduced hypergraph contains weighted hyperedges. 
\textsc{Relaxed-BIP} is our Binary Integer Programming formulation solved directly on the input hypergraph using \textsc{Gurobi} 11.0.3, with both \texttt{IntFeasTol} and \texttt{FeasibilityTol} set to $10^{-7}$ for higher numerical precision. All decision variables are initialized to zero prior to optimization. We evaluate it on both weighted and unweighted instances to provide a direct comparison to \ouralg~without reduction pre-processing.

\subparagraph{Setup and Reproducibility.} 
All experiments were run on a single core of a machine with an \texttt{AMD EPYC 7702P} CPU and 996\,GB RAM. The CPU has 64 physical (128 logical) cores, with clock speeds between 1.5 and 2\,GHz, a 64\,MiB L2 cache (0.5\,MiB per thread), and uses the~\texttt{x86\_64} architecture. The system runs \texttt{Ubuntu 20.04.1 LTS} with Linux kernel~\texttt{5.4.0-187-generic}.
We implemented \ouralg, and the competing algorithms \textsc{Trimmer} and \textsc{Relaxed-BIP} in C++14 and compiled them using gcc 11.4.0 with full optimization enabled (-O3 flag). In the study presented here, we test \ouralg~without label propagation and with label propagation using one iteration. We set the reduction threshold to~\numprint{1000} vertices. In our experiments, however, the hypergraph was often reduced well below this threshold, because the final round of reductions -- performed when the hypergraph still has more than~\numprint{1000} vertices -- can significantly reduce the hypergraph, often collapsing it to the fully reduced case (see Section~\ref{subsec:impact_red}). The code for \ouralg, as well as our implementation of \textsc{Trimmer}, will be publicly available on acceptance of the paper.   
\subparagraph{Real-World Instances.}  
To evaluate our algorithm, we use diverse hypergraphs from multiple domains. Our main test sets are the $M_{HG}$ dataset (488 medium-sized hypergraphs) \hbox{and the $L_{HG}$ dataset (94 large ones),} both used in prior work~\cite{DBLP:journals/corr/abs-2303-17679}. The hypergraphs originate from four well-established sources spanning three application domains: the ISPD98 VLSI Circuit Benchmark Suite~\cite{circuitbenchmark}, the DAC 2012 Routability-Driven Placement Contest~\cite{dacset}, the SuiteSparse Matrix Collection~\cite{sparsecollection}, and the International SAT Competition 2014~\cite{satbenchmark}. All hypergraphs are initially unweighted; we create weighted versions, by assigning each vertex and each hyperedge a random weight drawn uniformly from the range~$[1, 100]$. 

\subparagraph{Synthetic Instances.}  
We construct a dataset of~$(k, 2)$-core hypergraphs to evaluate performance on instances where the minimum cut~$\lambda$ is strictly smaller than the trivial cut~$\delta_{\omega}$. Inspired by \textsc{VieCut}~\cite{viecut}, we perform a~$(k, 2)$-core decomposition on the $L_{HG}$ instances by iteratively removing vertices with unweighted degree less than~$k$ and all hyperedges of size less than two. For each $L_{HG}$ hypergraph, we retain the~$(k, 2)$-core with the smallest~$k \geq 2$ where~$\lambda < \delta_{\omega}$. This yields 44 suitable~$(k, 2)$-core instances.

\subparagraph{Methodology.} We evaluate all algorithms under fixed resource limits: 2 hours runtime and 100 GB memory for $M_{HG}$ and 300 GB for $L_{HG}$ and~$(k, 2)$-core instances. Experiments were run using GNU Parallel~\cite{DBLP:journals/usenix-login/Tange11}, with 13 $M_{HG}$ instances and 4 $L_{HG}$ and~$(k, 2)$-core instances run concurrently.
An algorithm fails on an instance if it exceeds the time or memory limit before reporting a solution. \textsc{Relaxed-BIP} and~\ouralg~return the best feasible solution found so far if time runs out, but fail if memory is exceeded. We compare algorithms by runtime, cut value, and memory usage, aggregating results across instances with performance profiles~\cite{pp}. For each performance metric, the $x$-axis represents a factor $\tau \geq 1$, and the $y$-axis denotes the fraction of instances for which an algorithm’s performance is within a factor $\tau$ of the best-performing algorithm on that instance. More formally, let $\mathcal{A}$ be the set of all algorithms, $\mathcal{I}$ the set of instances, and $q_A(I)$ the quality of algorithm $A \in \mathcal{A}$ on instance $I \in \mathcal{I}$. For each algorithm $A$, we plot the fraction of instances $\frac{|\mathcal{I}_A(\tau)|}{|\mathcal{I}|}$ (y-axis) where $\mathcal{I}_A(\tau) := \set{I \in \mathcal{I} | q_A(I) \leq \tau \cdot min_{A' \in \mathcal{A}}q_{A'}(I)}$ \hbox{and $\tau$ is on the x-axis.
Performance profiles} accurately depict comparisons between algorithms even if some algorithm is unable to solve an instance: the curve for the corresponding algorithm enters into \hbox{the failed "F" region in the plot.} 

\subsection{Impact of Reductions}
\label{subsec:impact_red}
\begin{figure}[t]
    \centering
    \includegraphics[width=1\linewidth]{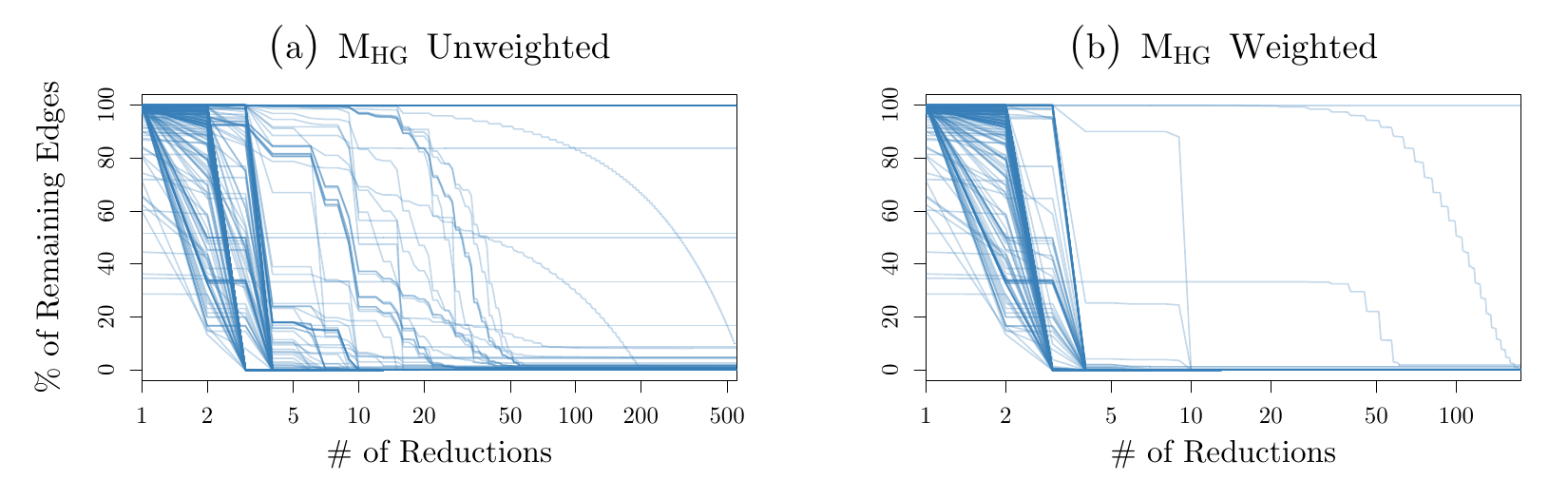}
    \caption{Reduction in hypergraph size over each \textbf{exact reduction} (no label propagation) on $M_{HG}$ instances. Each line shows one hypergraph. Reduction is measured as the percentage of remaining edges after each reduction, i.e., the reduced hypergraph size as a percentage of the input.}
    \label{fig:med-red-it0}
\end{figure}

To address \textbf{RQ1}, we evaluate the effectiveness of our reduction rules.  
Specifically, we measure the percentage of remaining hyperedges relative to the input -- i.e., the size of the reduced hypergraph as a percentage of the input hypergraph -- after each exact reduction, without the use of heuristic clustering.
Figure~\ref{fig:med-red-it0} demonstrates the effect of applying our reductions on the $M_{HG}$ instances, with each line representing a hypergraph instance. We observe that our exact reduction rules contract the hypergraph to a \emph{fully} reduced case -- i.e., a single vertex or zero hyperedges -- for the majority of both unweighted and weighted instances in fewer than five rounds. The hypergraph is fully reduced in %
85\% of unweighted and 95\% of weighted instances.  
Additional reductions continue to yield progress in the remaining harder instances. 
Similar patterns hold for the $L_{HG}$ instances, where the reduction rules yield fully reduced hypergraphs in 90\% of unweighted and 95\% of weighted instances.
These results demonstrate the strong practical effectiveness of our exact reductions on hypergraphs, even without the use of heuristic reduction via label propagation. Notably, for instances for which the reduction rules reduced the hypergraph to its fully reduced case, the minimum cut was already found, without running \textsc{Relaxed-BIP}, thus computing the exact minimum cut and improving the time and memory efficiency of \ouralg~in the majority of instances.

\subsection{Experiments on Real-World Hypergraphs}
\label{subsec:real}
\begin{figure}[t]
  \centering

  \begin{minipage}[t]{\linewidth}
    \centering
    \includegraphics[width=\linewidth]{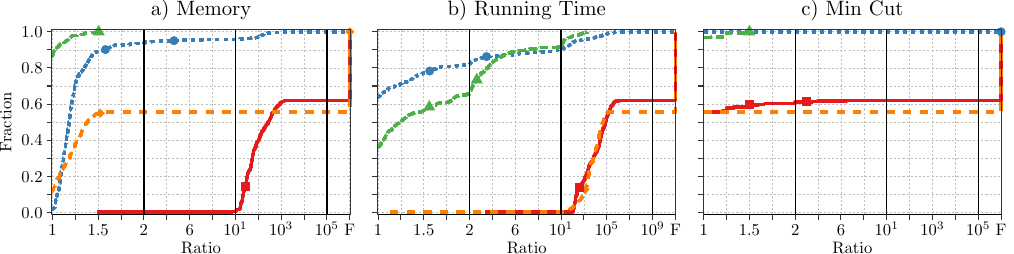}
    
    \vspace{0.5em}
    
    \includegraphics[width=0.7\linewidth]{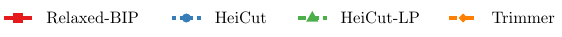}
    \subcaption{$M_{HG}$ Unweighted}
    \label{fig:medium-unweighted}
  \end{minipage}

  \vspace{1.5em}

  \begin{minipage}[t]{\linewidth}
    \centering
    \includegraphics[width=\linewidth]{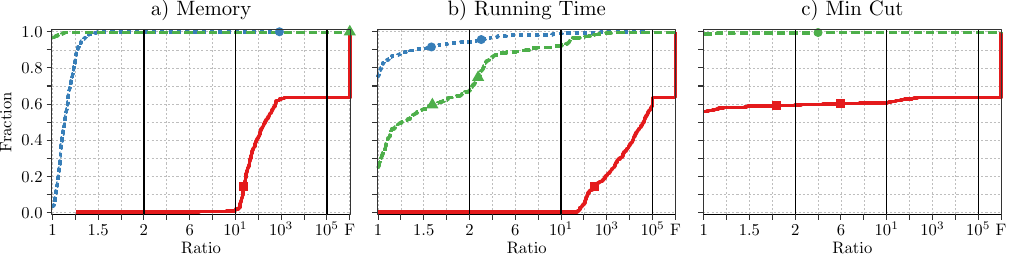}
    
    \vspace{0.5em}
    
    \includegraphics[width=0.5\linewidth]{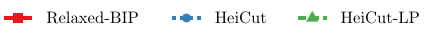}
    \subcaption{$M_{HG}$ Weighted}
    \label{fig:medium-weighted}
  \end{minipage}

  \caption{Performance profiles of \ouralg~with \textsc{Relaxed-BIP} and \textsc{Trimmer} on weighted and unweighted instances of the $M_{HG}$ dataset. \textsc{Trimmer} can only be executed on unweighted graphs.}
  \label{fig:medium-dataset}
\end{figure}
To answer \textbf{RQ2}, we compare \ouralg~to the baseline algorithms in terms of runtime and memory use on the $M_{HG}$ and $L_{HG}$ datasets. 
We also address \textbf{RQ3} by analyzing how the heuristic reduction via label propagation affects performance and accuracy.

Figure~\ref{fig:medium-dataset} shows performance profiles on the $M_{HG}$ dataset.
On unweighted instances, within the specified computational constraints, \ouralg~computed the best minimum cut value in all but one instance (\texttt{H20}), while \textsc{Trimmer} was only able to compute the minimum cut on approximately 55\% of instances, and \textsc{Relaxed-BIP} reported a cut on only approximately 60\% of instances, exceeding the memory limit on the remaining instances, as shown in Figure~\ref{fig:medium-unweighted}. On all instances on which \textsc{Trimmer} obtained the exact cut, \ouralg~was faster, and at least 1000 $\times$ faster in 85\% of such instances and more memory-efficient in 80\% of them. In approximately 50\% of such instances, \ouralg~used less than 40\% of \textsc{Trimmer}’s memory requirements (Figure~\ref{fig:medium-unweighted}). Additionally, our experimental results confirm the near-optimality of the LP relaxation of \textsc{Relaxed-BIP}: in all 145 instances where both \textsc{Trimmer} computed the exact cut and \textsc{Relaxed-BIP} completed within resource limits, \textsc{Relaxed-BIP} returned a solution matching the exact cut value.
Figure~\ref{fig:medium-weighted} shows that similar results hold on the weighted instances. %
 \ouralg~computed the best minimum cut on all weighted $M_{HG}$ instances, while \textsc{Relaxed-BIP} reported a cut on only 60\% of instances. 
 For all weighted and unweighted instances on which \textsc{Relaxed-BIP} computed a cut, \ouralg~was faster and more memory-efficient, being over 100 $\times$ faster and using less than 9\% of \textsc{Relaxed-BIP}’s memory requirement in 90\% of these instances. These results demonstrate the impact of our pre-processing reductions on the scalability of our algorithm.
\begin{figure}[t]
  \centering

  \begin{minipage}[t]{\linewidth}
    \centering
    \includegraphics[width=\linewidth]{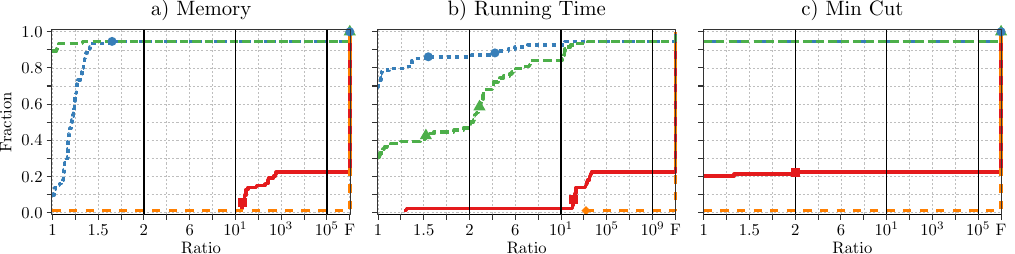}
    
    \vspace{0.5em}
    
    \includegraphics[width=0.7\linewidth]{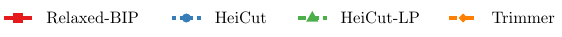}
    \subcaption{$L_{HG}$ Unweighted}
    \label{fig:large-unweighted}
  \end{minipage}

  \vspace{1.5em}

  \begin{minipage}[t]{\linewidth}
    \centering
    \includegraphics[width=\linewidth]{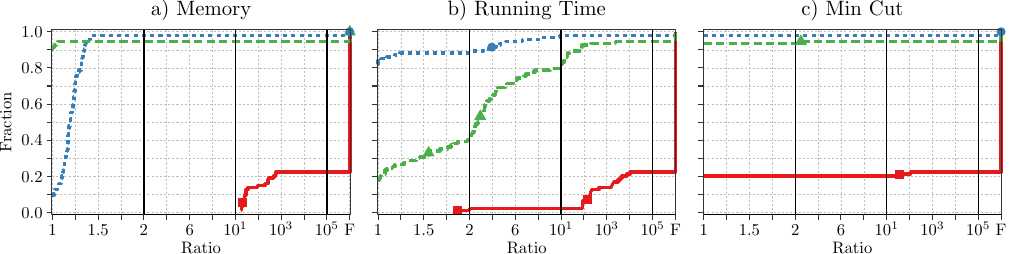}
    
    \vspace{0.5em}
    
    \includegraphics[width=0.5\linewidth]{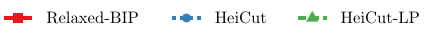}
    \subcaption{$L_{HG}$ Weighted}
    \label{fig:large-weighted}
  \end{minipage}

  \caption{Performance profiles of \ouralg~with \textsc{Relaxed-BIP} and \textsc{Trimmer} on weighted and unweighted instances of the $L_{HG}$ dataset. \textsc{Trimmer} can only be executed on unweighted graphs.}
  \label{fig:large-dataset}
\end{figure}

On the $L_{HG}$ dataset, \ouralg~computed the best minimum cut in approximately 95\% of instances, while \textsc{Trimmer} was able to compute the minimum cut in only one unweighted instance, and \textsc{Relaxed-BIP} reported a cut in only about 20\% of weighted and unweighted instances, as shown in Figure~\ref{fig:large-dataset}, failing in the majority of instances by exceeding the memory limit.~\ouralg~outperformed \textsc{Relaxed-BIP} in both time and memory efficiency on all weighted and unweighted $L_{HG}$ instances, being more than 100 $\times$ faster and using less than 9\% memory in over 95\% of all the instances that \textsc{Relaxed-BIP} successfully ran on.

To address \textbf{RQ3}, we compare \ouralg~with and without label propagation. On real-world datasets, despite using a heuristic reduction, \ouralg~with label propagation still computed the best minimum cut on over 95\% of instances, as seen in Figure~\ref{fig:medium-dataset} and Figure~\ref{fig:large-dataset}. \ouralg~with label propagation also computed a minimum cut on the single $M_{HG}$ instance on which \ouralg~without label propagation could not. 
The use of label propagation improves the memory efficiency of \ouralg~with minimal compromise in the quality, though at the cost of runtime.
Although it accelerates the reduction step, the runtime cost of computing label propagation outweighs these gains, making \ouralg~without label propagation faster in most instances.
While applying label propagation reduces memory consumption, \ouralg~is very memory-efficient even without it -- the memory usage stays below 200~MB for $M_{HG}$ and 2~GB for $L_{HG}$ instances.

\subsection{ Experiments on Synthetic Hypergraphs}
\label{subsec:syn}
\begin{figure}[t]
  \centering

  \begin{minipage}[t]{\linewidth}
    \centering
    \includegraphics[width=\linewidth]{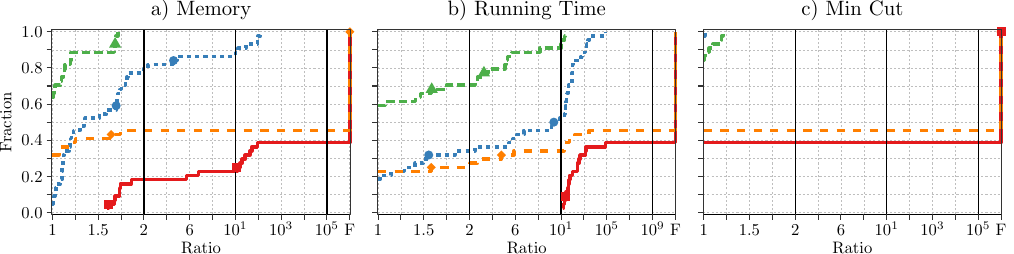}
    
    \vspace{0.5em}
    
    \includegraphics[width=0.7\linewidth]{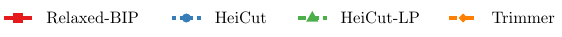}
    \subcaption{$(k, 2)$-Core Unweighted}
    \label{fig:kcore-unweighted}
  \end{minipage}

  \vspace{1.5em}

  \begin{minipage}[t]{\linewidth}
    \centering
    \includegraphics[width=\linewidth]{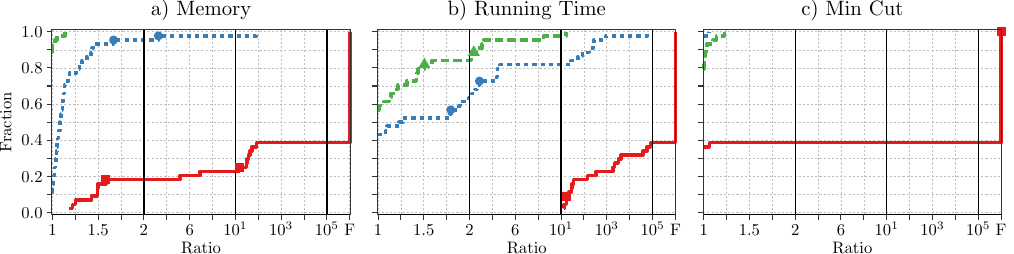}
    
    \vspace{0.5em}
    
    \includegraphics[width=0.5\linewidth]{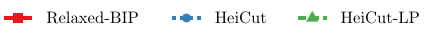}
    \subcaption{$(k, 2)$-Core Weighted}
    \label{fig:kcore-weighted}
  \end{minipage}

  \caption{Performance profiles of \ouralg~with \textsc{Relaxed-BIP} and \textsc{Trimmer} on weighted and unweighted instances of the~$(k, 2)$-core dataset. \textsc{Trimmer} can only be \hbox{executed on unweighted graphs.}}
  \label{fig:kcore-dataset}
\end{figure}
To further address \textbf{RQ2} and \textbf{RQ3}, we test the algorithms on synthetic~$(k, 2)$-core hypergraphs, which are constructed such that the minimum cut value $\lambda$ is strictly smaller than the trivial bound $\delta_\omega$.
Our experiments on~$(k, 2)$-core hypergraphs demonstrate that \ouralg~outperforms \textsc{Trimmer} and \textsc{Relaxed-BIP} even on harder instances where the minimum cut is not the trivial cut. \ouralg~computes the best cut on all unweighted and weighted instances, while \textsc{Trimmer} computes the exact cut on only 50\% of unweighted instances, and \textsc{Relaxed-BIP} computes a cut on only 38\% of unweighted and weighted instances, as shown in Figures~\ref{fig:kcore-unweighted} and~\ref{fig:kcore-weighted}. \ouralg~is faster and more memory-efficient than \textsc{Relaxed-BIP} in all successful unweighted and weighted instances. Interestingly, while~\ouralg~is faster without label propagation on real-world instances, \ouralg~with label propagation is faster than \ouralg~without label propagation on~$(k, 2)$-core instances, while still computing the best minimum cut in 80\% of instances, as shown in Figure~\ref{fig:kcore-unweighted} and Figure~\ref{fig:kcore-weighted}. \ouralg~with label propagation achieves the best runtime in 60\% of instances, and uses the least memory in 60\% of unweighted and 90\% of weighted instances. These results demonstrate the effectiveness of using label propagation along with the reductions. %

\section{Conclusion}
We presented \ouralg, a scalable algorithm for computing near-optimal minimum cuts in both weighted and unweighted hypergraphs. Our approach combines novel exact reduction rules with generalizations of proven techniques from graph algorithms, enabling aggressive size reductions while rigorously preserving correctness. After shrinking the hypergraph, we apply a Binary Integer Programming formulation that computes a near-exact minimum cut. 
Extensive experiments on more than 500 real-world and synthetic instances demonstrate the effectiveness of our approach: in over 85\% of instances, the reduction rules alone suffice to identify the exact minimum cut, eliminating the need to run the relaxed BIP solver. \ouralg~scaled to twice as many instances as the state-of-the-art hypergraph minimum cut algorithm, and was up to five orders of magnitude faster than it. \ouralg~thus establishes a new benchmark for hypergraph minimum cut computation, enabling the novel use of a hypergraph minimum cut solver as a practical subroutine in large-scale optimization problems. We further introduced a new benchmark of~$(k, 2)$-core hypergraphs designed to evaluate solver performance on non-trivial cuts, offering a valuable test set for future research. In future work, we aim to parallelize our reduction routines, and consider extensions to \hbox{related problems.}

\bibliography{compact_arxiv}

\begin{thebibliography}{10}

\bibitem{circuitbenchmark}
Charles~J. Alpert.
\newblock {T}he {ISPD98} {C}ircuit {B}enchmark {S}uite.
\newblock In Majid Sarrafzadeh, editor, {\em Proceedings of the 1998
  International Symposium on Physical Design, {ISPD} 1998, Monterey, CA, USA,
  April 6-8, 1998}, pages 80--85, 1998.
\newblock \href {https://doi.org/10.1145/274535.274546}
  {\path{doi:10.1145/274535.274546}}.

\bibitem{satbenchmark}
Anton Belov, Daniel Diepold, Marijn Heule, and Matti J\"arvisalo.
\newblock {T}he {SAT} {C}ompetition 2014.
\newblock \url{http://www.satcompetition.org/2014/index.shtml}, 2014.

\bibitem{doi:10.1126/science.aad9029}
Austin~R. Benson, David~F. Gleich, and Jure Leskovec.
\newblock {H}igher-order {O}rganization of {C}omplex {N}etworks.
\newblock {\em CoRR}, abs/1612.08447, 2016.
\newblock URL: \url{http://arxiv.org/abs/1612.08447}, \href
  {https://arxiv.org/abs/1612.08447} {\path{arXiv:1612.08447}}.

\bibitem{cai2005community}
Deng Cai, Zheng Shao, Xiaofei He, Xifeng Yan, and Jiawei Han.
\newblock {M}ining {H}idden {C}ommunity in {H}eterogeneous {S}ocial {N}etworks.
\newblock In Jafar Adibi, Marko Grobelnik, Dunja Mladenic, and Patrick Pantel,
  editors, {\em Proceedings of the 3rd international workshop on Link
  discovery, LinkKDD 2005, Chicago, Illinois, USA, August 21-25, 2005}, pages
  58--65, 2005.
\newblock \href {https://doi.org/10.1145/1134271.1134280}
  {\path{doi:10.1145/1134271.1134280}}.

\bibitem{DBLP:journals/tpds/CatalyurekA99}
{\"{U}}mit~V. {\c{C}}ataly{\"{u}}rek and Cevdet Aykanat.
\newblock {H}ypergraph-{P}artitioning-based {D}ecomposition for {P}arallel
  {S}parse-matrix {V}ector {M}ultiplication.
\newblock {\em {IEEE} Trans. Parallel Distributed Syst.}, 10(7):673--693, 1999.
\newblock \href {https://doi.org/10.1109/71.780863}
  {\path{doi:10.1109/71.780863}}.

\bibitem{chekurihypergraphmincut}
Chandra Chekuri and Chao Xu.
\newblock {C}omputing {M}inimum {C}uts in {H}ypergraphs.
\newblock In Philip~N. Klein, editor, {\em Proceedings of the Twenty-Eighth
  Annual {ACM-SIAM} Symposium on Discrete Algorithms, {SODA} 2017, Barcelona,
  Spain, Hotel Porta Fira, January 16-19}, pages 1085--1100, 2017.
\newblock \href {https://doi.org/10.1137/1.9781611974782.70}
  {\path{doi:10.1137/1.9781611974782.70}}.

\bibitem{sparsecollection}
Timothy~A. Davis and Yifan Hu.
\newblock {T}he {U}niversity of {F}lorida {S}parse {M}atrix {C}ollection.
\newblock {\em {ACM} Trans. Math. Softw.}, 38(1):1:1--1:25, 2011.
\newblock \href {https://doi.org/10.1145/2049662.2049663}
  {\path{doi:10.1145/2049662.2049663}}.

\bibitem{pp}
Elizabeth~D. Dolan and Jorge~J. Mor{\'{e}}.
\newblock {B}enchmarking {O}ptimization {S}oftware {W}ith {P}erformance
  {P}rofiles.
\newblock {\em Math. Program.}, 91(2):201--213, 2002.
\newblock URL: \url{https://doi.org/10.1007/s101070100263}, \href
  {https://doi.org/10.1007/S101070100263} {\path{doi:10.1007/S101070100263}}.

\bibitem{DBLP:journals/corr/abs-2303-17679}
Lars Gottesb{\"{u}}ren, Tobias Heuer, Nikolai Maas, Peter Sanders, and
  Sebastian Schlag.
\newblock {S}calable {H}igh-quality {H}ypergraph {P}artitioning.
\newblock {\em {ACM} Trans. Algorithms}, 20(1):9:1--9:54, 2024.
\newblock \href {https://doi.org/10.1145/3626527} {\path{doi:10.1145/3626527}}.

\bibitem{mtkahypar}
Lars Gottesb{\"{u}}ren, Tobias Heuer, Peter Sanders, and Sebastian Schlag.
\newblock {S}calable {S}hared-memory {H}ypergraph {P}artitioning.
\newblock In Martin Farach{-}Colton and Sabine Storandt, editors, {\em
  Proceedings of the Symposium on Algorithm Engineering and Experiments,
  {ALENEX} 2021, Virtual Conference, January 10-11, 2021}, pages 16--30, 2021.
\newblock \href {https://doi.org/10.1137/1.9781611976472.2}
  {\path{doi:10.1137/1.9781611976472.2}}.

\bibitem{haomincut1992}
Jianxiu Hao and James~B. Orlin.
\newblock {A} {F}aster {A}lgorithm for {F}inding the {M}inimum {C}ut in a
  {G}raph.
\newblock In Greg~N. Frederickson, editor, {\em Proceedings of the Third Annual
  {ACM/SIGACT-SIAM} Symposium on Discrete Algorithms, 27-29 January 1992,
  Orlando, Florida, {USA}}, pages 165--174, 1992.
\newblock URL: \url{http://dl.acm.org/citation.cfm?id=139404.139439}.

\bibitem{HARTUV2000175}
Erez Hartuv and Ron Shamir.
\newblock {A} {C}lustering {A}lgorithm {B}ased on {G}raph {C}onnectivity.
\newblock {\em Inf. Process. Lett.}, 76(4-6):175--181, 2000.
\newblock \href {https://doi.org/10.1016/S0020-0190(00)00142-3}
  {\path{doi:10.1016/S0020-0190(00)00142-3}}.

\bibitem{Henne2015_1000063440}
Vitali Henne.
\newblock {L}abel {P}ropagation for {H}ypergraph {P}artitioning.
\newblock Master's thesis, 2015.
\newblock \href {https://doi.org/10.5445/IR/1000063440}
  {\path{doi:10.5445/IR/1000063440}}.

\bibitem{viecut}
Monika Henzinger, Alexander Noe, and Christian Schulz.
\newblock {S}hared-memory {E}xact {M}inimum {C}uts.
\newblock In {\em 2019 {IEEE} International Parallel and Distributed Processing
  Symposium, {IPDPS} 2019, Rio de Janeiro, Brazil, May 20-24, 2019}, pages
  13--22, 2019.
\newblock \href {https://doi.org/10.1109/IPDPS.2019.00013}
  {\path{doi:10.1109/IPDPS.2019.00013}}.

\bibitem{henzingerflowedgecon2020}
Monika Henzinger, Satish Rao, and Di~Wang.
\newblock {L}ocal {F}low {P}artitioning for {F}aster {E}dge {C}onnectivity.
\newblock In Philip~N. Klein, editor, {\em Proceedings of the Twenty-Eighth
  Annual {ACM-SIAM} Symposium on Discrete Algorithms, {SODA} 2017, Barcelona,
  Spain, Hotel Porta Fira, January 16-19}, pages 1919--1938, 2017.
\newblock \href {https://doi.org/10.1137/1.9781611974782.125}
  {\path{doi:10.1137/1.9781611974782.125}}.

\bibitem{kargernetworkreliability}
David~R. Karger.
\newblock {A} {R}andomized {F}ully {P}olynomial {T}ime {A}pproximation {S}cheme
  for the {A}ll {T}erminal {N}etwork {R}eliability {P}roblem.
\newblock In Frank~Thomson Leighton and Allan Borodin, editors, {\em
  Proceedings of the Twenty-Seventh Annual {ACM} Symposium on Theory of
  Computing, 29 May-1 June 1995, Las Vegas, Nevada, {USA}}, pages 11--17, 1995.
\newblock \href {https://doi.org/10.1145/225058.225069}
  {\path{doi:10.1145/225058.225069}}.

\bibitem{karger2000mincut}
David~R. Karger.
\newblock {M}inimum {C}uts in {N}ear-linear {T}ime.
\newblock In Gary~L. Miller, editor, {\em Proceedings of the Twenty-Eighth
  Annual {ACM} Symposium on the Theory of Computing, Philadelphia,
  Pennsylvania, USA, May 22-24, 1996}, pages 56--63, 1996.
\newblock \href {https://doi.org/10.1145/237814.237829}
  {\path{doi:10.1145/237814.237829}}.

\bibitem{kargermincut1996}
David~R. Karger and Clifford Stein.
\newblock {A} {N}ew {A}pproach to the {M}inimum {C}ut {P}roblem.
\newblock {\em J. {ACM}}, 43(4):601--640, 1996.
\newblock \href {https://doi.org/10.1145/234533.234534}
  {\path{doi:10.1145/234533.234534}}.

\bibitem{DBLP:journals/tvlsi/KarypisAKS99}
George Karypis, Rajat Aggarwal, Vipin Kumar, and Shashi Shekhar.
\newblock {M}ultilevel {H}ypergraph {P}artitioning: {A}pplications in {VLSI}
  {D}omain.
\newblock {\em {IEEE} Trans. Very Large Scale Integr. Syst.}, 7(1):69--79,
  1999.
\newblock \href {https://doi.org/10.1109/92.748202}
  {\path{doi:10.1109/92.748202}}.

\bibitem{Klimmek1996}
Regina Klimmek and Frank Wagner.
\newblock {A} {S}imple {H}ypergraph {M}in {C}ut {A}lgorithm.
\newblock Technical Report B 96-02, Bericht FU Berlin Fachbereich Mathematik
  und Informatik, 1996.
\newblock URL: \url{http://edocs.fu-berlin.de/docs/servlets/MCRFileNodeServlet/
  FUDOCS_derivate_000000000297/1996_02.pdf}.

\bibitem{DBLP:conf/icdcn/KothapalliPS13}
Kishore Kothapalli, Sriram~V. Pemmaraju, and Vivek Sardeshmukh.
\newblock {O}n the {A}nalysis of a {L}abel {P}ropagation {A}lgorithm for
  {C}ommunity {D}etection.
\newblock In Davide Frey, Michel Raynal, Saswati Sarkar, Rudrapatna~K.
  Shyamasundar, and Prasun Sinha, editors, {\em Distributed Computing and
  Networking, 14th International Conference, {ICDCN} 2013, Mumbai, India,
  January 3-6, 2013. Proceedings}, volume 7730 of {\em Lecture Notes in
  Computer Science}, pages 255--269, 2013.
\newblock \href {https://doi.org/10.1007/978-3-642-35668-1\_18}
  {\path{doi:10.1007/978-3-642-35668-1\_18}}.

\bibitem{KrishnamurthyVLSI}
Krishnamurthy.
\newblock {A}n {I}mproved {M}in-cut {A}lgonthm for {P}artitioning {V}{L}{S}{I}
  {N}etworks.
\newblock {\em IEEE Transactions on Computers}, C-33(5):438--446, 1984.
\newblock \href {https://doi.org/10.1109/TC.1984.1676460}
  {\path{doi:10.1109/TC.1984.1676460}}.

\bibitem{Lengauer1990}
Thomas Lengauer.
\newblock {\em {G}raph {A}lgorithms}, pages 47--135.
\newblock Vieweg+Teubner Verlag, Wiesbaden, 1990.
\newblock \href {https://doi.org/10.1007/978-3-322-92106-2_3}
  {\path{doi:10.1007/978-3-322-92106-2_3}}.

\bibitem{makwongmincut2000}
Wai{-}Kei Mak and D.~F. Wong.
\newblock {A} {F}ast {H}ypergraph {M}in-cut {A}lgorithm for {C}ircuit
  {P}artitioning.
\newblock {\em Integr.}, 30(1):1--11, 2000.
\newblock \href {https://doi.org/10.1016/S0167-9260(00)00008-0}
  {\path{doi:10.1016/S0167-9260(00)00008-0}}.

\bibitem{nagamochiedgecon1992}
Hiroshi Nagamochi and Toshihide Ibaraki.
\newblock {C}omputing {E}dge-connectivity in {M}ultigraphs and {C}apacitated
  {G}raphs.
\newblock {\em {SIAM} J. Discret. Math.}, 5(1):54--66, 1992.
\newblock \href {https://doi.org/10.1137/0405004} {\path{doi:10.1137/0405004}}.

\bibitem{nagamochimincut1994}
Hiroshi Nagamochi, Tadashi Ono, and Toshihide Ibaraki.
\newblock {I}mplementing an {E}fficient {M}inimum {C}apacity {C}ut {A}lgorithm.
\newblock {\em Math. Program.}, 67:325--341, 1994.
\newblock \href {https://doi.org/10.1007/BF01582226}
  {\path{doi:10.1007/BF01582226}}.

\bibitem{Padberg1990}
M.~Padberg and G.~Rinaldi.
\newblock {A}n {E}fficient {A}lgorithm for the {M}inimum {C}apacity {C}ut
  {P}roblem.
\newblock {\em Mathematical Programming}, 47(1):19--36, May 1990.
\newblock \href {https://doi.org/10.1007/BF01580850}
  {\path{doi:10.1007/BF01580850}}.

\bibitem{padbergtsp}
Manfred Padberg and Giovanni Rinaldi.
\newblock {A} {B}ranch-and-cut {A}lgorithm for the {R}esolution of
  {L}arge-scale {S}ymmetric {T}raveling {S}alesman {P}roblems.
\newblock {\em {SIAM} Rev.}, 33(1):60--100, 1991.
\newblock \href {https://doi.org/10.1137/1033004} {\path{doi:10.1137/1033004}}.

\bibitem{Queyranne1998}
Maurice Queyranne.
\newblock {M}inimizing {S}ymmetric {S}ubmodular {F}unctions.
\newblock {\em Math. Program.}, 82:3--12, 1998.
\newblock \href {https://doi.org/10.1007/BF01585863}
  {\path{doi:10.1007/BF01585863}}.

\bibitem{Raghavan_2007}
Usha~Nandini Raghavan, R{\'{e}}ka Albert, and Soundar Kumara.
\newblock {N}ear {L}inear {T}ime {A}lgorithm to {D}etect {C}ommunity
  {S}tructures in {L}arge-scale {N}etworks.
\newblock {\em Physical Review E}, 76(3), 2007.
\newblock URL: \url{https://doi.org/10.1103%2Fphysreve.76.036106}, \href
  {https://doi.org/10.1103/physreve.76.036106}
  {\path{doi:10.1103/physreve.76.036106}}.

\bibitem{Ramanathannetworkreliability}
Aparna Ramanathan and Charles~J. Colbourn.
\newblock {C}ounting {A}lmost {M}inimum {C}utsets {W}ith {R}eliability
  {A}pplications.
\newblock {\em Math. Program.}, 39(3):253--261, 1987.
\newblock \href {https://doi.org/10.1007/BF02592076}
  {\path{doi:10.1007/BF02592076}}.

\bibitem{DBLP:journals/usenix-login/Tange11}
Ole Tange.
\newblock {GNU} {P}arallel: {T}he {C}ommand-line {P}ower {T}ool.
\newblock {\em login Usenix Mag.}, 36(1), 2011.
\newblock URL:
  \url{https://www.usenix.org/publications/login/february-2011-volume-36-number-1/gnu-parallel-command-line-power-tool}.

\bibitem{dacset}
Natarajan Viswanathan, Charles~J. Alpert, Cliff C.~N. Sze, Zhuo Li, and
  Yaoguang Wei.
\newblock {T}he {DAC} 2012 {R}outability-driven {P}lacement {C}ontest and
  {B}enchmark {S}uite.
\newblock In Patrick Groeneveld, Donatella Sciuto, and Soha Hassoun, editors,
  {\em The 49th Annual Design Automation Conference 2012, {DAC} '12, San
  Francisco, CA, USA, June 3-7, 2012}, pages 774--782, 2012.
\newblock \href {https://doi.org/10.1145/2228360.2228500}
  {\path{doi:10.1145/2228360.2228500}}.

\bibitem{wucutclustering1993}
Z.~Wu and R.~Leahy.
\newblock {A}n {O}ptimal {G}raph {T}heoretic {A}pproach to {D}ata {C}lustering:
  {T}heory and {I}ts {A}pplication to {I}mage {S}egmentation.
\newblock {\em IEEE Transactions on Pattern Analysis and Machine Intelligence},
  15(11):1101--1113, 1993.
\newblock \href {https://doi.org/10.1109/34.244673}
  {\path{doi:10.1109/34.244673}}.

\end{thebibliography}

\appendix
\section{Description and Implementation of the \textsc{Trimmer} Algorithm}
\label{appendix:trimmer}
\begin{algorithm}[t]
\caption{\textsc{Trimmer}: Minimum Cut of Unweighted Hypergraphs by Chekuri et. al~\cite{chekurihypergraphmincut}.}
\label{alg:trimmer}
\begin{algorithmic}
\State \textbf{Input:} Hypergraph~$H = (V, E)$
\State \textbf{Output:} Minimum cut value~$\lambda$

\State $o \gets$ \Call{ComputeHeadOrdering}{$H$} \Comment{\textcolor{gray}{via MA vertex ordering}}
\For{\textbf{each} hyperedge~$e$ in order~$o$}
    \For{\textbf{each} pin~$v \in e$ \textbf{where} $v$ is not the head of~$e$}
        \State Add~$e$ to backward edges of~$v$
    \EndFor
\EndFor

\State $k \gets 2$
\While{\textbf{true}}
    \State $H_k \gets$ \Call{ConstructTrimmedCertificate}{$H$, $k$} \Comment{\textcolor{gray}{using backward edges}}
    \State $\lambda \gets$ \Call{ComputeMinimumCut}{$H_k$} \Comment{\textcolor{gray}{exact solver}}
    \If{$k > \lambda$}
        \State \textbf{break}
    \EndIf
    \State $k \gets 2 \cdot k$
\EndWhile

\State \Return $\lambda$
\end{algorithmic}
\end{algorithm}
The \textsc{Trimmer} algorithm by Chekuri et al.~\cite{chekurihypergraphmincut} is a sparsification-based method for computing the minimum cut in unweighted hypergraphs. It constructs a sequence of sparsified hypergraphs, known as \emph{$k$-trimmed certificates}, each of which preserves all local connectivities up to a threshold~$k$. These certificates contain only~$O(kn)$ hyperedges and are passed to an exact minimum cut solver. The algorithm iteratively doubles~$k$ and computes the minimum cut of each certificate until the cut value of the original hypergraph is recovered. For further details and proofs of correctness, we refer the reader to the original work~\cite{chekurihypergraphmincut}.

\subparagraph{Implementation Details.}
We implement \textsc{Trimmer} as described by the authors as follows. To construct a $k$-trimmed certificate~$H_k$, the algorithm first computes a \emph{MA ordering}~\cite{nagamochiedgecon1992} of the vertices in~$H$, starting from a random vertex. 
In this ordering, for unweighted hypergraphs, each subsequent vertex~$v_i$ is selected to maximize the number of incident hyperedges connecting it to previously ordered vertices~$v_1, \dots, v_{i-1}$. 
This can be computed efficiently using a bucket priority queue. Next, we define the \emph{head} of each hyperedge~$e$ as its first pin in the MA ordering and sort hyperedges by the position of their heads, breaking ties using their original order. This produces the \emph{hyperedge head ordering}.

Using this hyperedge head ordering, we precompute, for each vertex~$v$, its list of \emph{backward edges}—hyperedges containing~$v$ for which~$v$ is not the head. These lists, sorted by the hyperedge head ordering, enable efficient construction of~$H_k$ for any $k$: for each vertex~$v$, we retain up to~$k$ of its backward edges in~$H_k$.

The algorithm proceeds in iterations, starting with~$k = 2$. In each iteration, we construct~$H_k$ and compute its minimum cut using an exact solver (e.g., Klimmek and Wagner~\cite{Klimmek1996}, Mak and Wong~\cite{makwongmincut2000}, or Queyranne~\cite{Queyranne1998}). If the computed cut value is strictly less than~$k$, then~$\lambda(H) = \lambda(H_k)$ and the algorithm terminates. Otherwise,~$k$ is doubled and the process continues. A pseudocode of our implementation is provided in Algorithm~\ref{alg:trimmer}.
\end{document}